\documentclass[letterpaper, 10 pt, conference]{ieeeconf}  

\IEEEoverridecommandlockouts                              
\usepackage{graphics} 
\usepackage{times} 
\usepackage{amsmath, amsfonts} 
\usepackage{amssymb}  
\usepackage{url}
\usepackage{cite}
\usepackage{hyperref}

\usepackage{algorithmic}
\usepackage{graphicx}
\usepackage{textcomp}
\usepackage{xcolor}

\title{\LARGE \bf
Adaptive Reconstruction of Nonlinear Systems States via DREM with Perturbation Annihilation
}

\author{Anton Glushchenko, \textit{Member, IEEE} and Konstantin Lastochkin
\thanks{A. Glushchenko is with V.A. Trapeznikov Institute of Control Sciences of Russian Academy of Sciences, Moscow, Russia
        {\tt\small aiglush@ipu.ru}}%
\thanks{K. Lastochkin is with V.A. Trapeznikov Institute of Control Sciences of Russian Academy of Sciences, Moscow, Russia 
        {\tt\small lastconst@ipu.ru}}%
}

\begin{document}

\maketitle
\thispagestyle{empty}
\pagestyle{empty}

\begin{abstract}
A new adaptive observer is proposed for a certain class of nonlinear systems with bounded unknown input and parametric uncertainty. Unlike most existing solutions, the proposed approach ensures {asymptotic} convergence of the unknown parameters, state and perturbation estimates to an arbitrarily small neighborhood of the equilibrium point. {The solution is based on the novel augmentation of a high-gain observer with the dynamic regressor extension and mixing (DREM) procedure enhanced with a perturbation annihilation algorithm.} The aforementioned properties of the proposed solution are verified via numerical experiments.
\end{abstract}

\section{Introduction}
{In technical systems only regulated output is usually available for measurement.} However, measurement or estimation {of all system states} is useful to improve the control quality (by feeding back the obtained estimates) and is necessary to ensure process safety, robustness and system {reliability} (by fault tolerance control and fault estimation). Starting with the studies by D. Luenberger \cite{b1} and R. Kalman \cite{b2}, various methods of state reconstruction for linear and nonlinear systems have been actively developed (see a recent review \cite{b3}). {If the system parameters are unknown, then the state reconstruction is augmented with parameter identification. Such observers with simultaneous estimation of state and unknown parameters are called adaptive.}

The first adaptive observer was proposed by R. Carrol and D. Lindorf \cite{b4}. After a number of generalizations of their result, an observer was proposed in \cite{b5} that, unlike \cite{b4}, ensures an exponential convergence rate of both parametric and state estimation errors. The main assumptions of \cite{b4, b5} are that the requirement of the regressor persistent excitation is met, and the system is represented in the observer canonical form. These assumptions were substituted (but not relaxed) with the output matching condition in \cite{b6}, and later -- with the extended matching condition \cite{b7}. Starting from \cite{b8,b9,b10}, the interest of researchers has shifted to the class of nonlinear systems which, using the output injection, can be transformed into a form that is equivalent to the observer canonical one for linear systems \cite{b10,b11}. Then, in \cite{b12}, a unified form of the nonlinear systems adaptive observer was proposed, which provides asymptotic estimation of the state and parameter vectors under some conditions that are similar to the ones of the passification problem solvability. In \cite{b13}, a more general form of the adaptive observers for a class of uniformly observable nonlinear systems with a single output was proposed. Having injected an additional signal, R. Marino and P. Tomei generalized the results of \cite{b5} to nonlinear systems \cite{b14}, providing an exponential convergence rate. In \cite{b15}, an observer was proposed that allows one to obtain state estimate via algebraic rather than differential equation. As for the external perturbations, most of the above-considered algorithms are input to state stable (ISS) with respect to them \cite{b16}. However, the uniform upper bound (UUB) of parameter and state estimation errors can be arbitrarily large especially for high-amplitude disturbances. To overcome this drawback, adaptive observers were proposed in \cite{b17,b18} on the basis of the internal model principle to parameterize and estimate the unknown perturbations. At the same time, the known perturbation model is a restrictive assumption for practical scenarios. To overcome this disadvantage, in \cite{b19,b20,b21} the adaptive observers were proposed to be augmented with the sliding-mode-based perturbation observers. As a result, hybrid observers were derived with improved robustness to external perturbations in the sense of smallness of the state estimation steady-state error compared to pure adaptive observers \cite{b4,b5,b6,b7,b8,b9,b10,b11,b12,b13,b14,b15,b16}. However, the following main drawback of the solutions \cite{b19,b20,b21} should be noticed:

\textbf{MD}. the steady-state error (UUB) of the unknown parameters and system state estimation cannot be made arbitrarily small regardless of the choice of the adaptive observer parameters (see Theorem 2 in \cite{b19}, bounds (16) and (22) in \cite{b20}, and Theorem 4 in \cite{b21}).

This drawback is caused by the fact that the estimation laws used to identify the unknown parameters are robust to perturbations in the sense of uniform ultimate boundedness (UUB), but they do not guarantee asymptotic convergence of the unknown parameter estimates to their true values in case of even small perturbations. In this study, this drawback is overcome by application of a new estimation law \cite{b22}, which ensures asymptotic convergence of the parametric error to an arbitrarily small neighborhood of zero even if the system is affected by an unknown but bounded perturbation.

The proposed adaptive observer with asymptotic estimation of unknown parameters for perturbed systems is implemented in the following way. A high-gain observer from \cite{b23} is applied to estimate a bounded external perturbation with an arbitrary accuracy. To identify the unknown parameters, a linear regression equation is parameterized, which is used to design the estimation law \cite{b22}. The convergence of the unknown parameters estimates to an arbitrarily small neighborhood of the true values is achieved when: {\it i}) a condition similar to the regressor persistent excitation one is met, {\it ii}) at least one element of the {parameterized} regressor is independent from the perturbation. The magnitude of the steady-state error is inversely proportional to a certain arbitrary parameter of the above-mentioned law. Using the second Lyapunov method, it is proved that, if the parametric convergence conditions are satisfied, the proposed observer, unlike the solutions \cite{b19,b20,b21}, guarantees asymptotic convergence of the state reconstruction error to an arbitrarily small neighborhood of the equilibrium point.

The remainder of the paper is organized as follows. The problem statement is given in Section {II}. Section {III} is to elucidate the proposed adaptive observer and investigate its properties. Simulation results are shown in Section {IV}.

\textbf{Notation.} Further the following notation is used: $\left| . \right|$ is the absolute value, $\left\| . \right\|$ is the suitable norm of $(.)$, ${I_{n \times n}}=I_{n}$ is an identity $n \times n$ matrix, ${0_{n \times n}}$ is a zero $n \times n$ matrix, $0_{n}$ stands for a zero vector of length $n$, ${\rm{det}}\{.\}$ stands for a matrix determinant, ${\rm{adj}}\{.\}$ represents an adjoint matrix. We also use the fact that for all (possibly singular) ${n \times n}$ matrices $M$ the following holds: ${\rm{adj}} \{M\} M = {\rm{det}} \{M\}I_{n \times n}$.

\section{Problem Statement}
The following class of nonlinear systems is considered:
\begin{equation}\label{eq1}
\begin{array}{l}
\dot x\left( t \right) = Ax\left( t \right) + \phi \left( {y{\rm{,\;}}u} \right) + G\left( {y{\rm{,\;}}u} \right)\theta  + D\delta \left( t \right){\rm{,}}\\
y\left( t \right) = Cx\left( t \right){\rm{,\;}}x\left( {{t_0}} \right) = {x_0}{\rm{,}}
\end{array}    
\end{equation}
where $x\left( t \right) \in {\mathbb{R}^n}{\rm{,\;}}y\left( t \right) \in {\mathbb{R}^p}{\rm{,\;}}u\left( t \right) \in {D_u} \subset {\mathbb{R}^m}$ are state, output and control signals, respectively, $\theta  \in {D_\theta } \subset {\mathbb{R}^q}$ stands for an unknown parameters vector, $\delta \left( t \right) \in {\mathbb{R}^s}$ denotes an exogenous perturbation, the mappings $\phi {\rm{:\;}}{\mathbb{R}^p} \times {\mathbb{R}^m} \mapsto {\mathbb{R}^n}$ and $G{\rm{:\;}}{\mathbb{R}^p} \times {\mathbb{R}^m} \mapsto {\mathbb{R}^{n \times q}}$ are known and ensure uniqueness and existence of solutions to system (1). The matrices \linebreak $A \in {\mathbb{R}^{n \times n}}{\rm{,\;}}{D}  \in {\mathbb{R}^{n \times s}}{\rm{,\;}}{{C}}  \in {\mathbb{R}^{p \times n}}$ are known, and the pair $\left( {A{\rm{,\;}}C} \right)$ is detectable. The following assumptions are adopted with respect to the signals $\delta \left( t \right){\rm{,\;}}G\left( {y{\rm{,\;}}u} \right)$ and structure of the linear part of the system (1).

\emph{\textbf{Assumption 1.} There exist ${\delta _{{\rm{max}}}} = \mathop {{\rm{sup}}}\limits_t \left\| {\delta \left( t \right)} \right\|$ and ${\dot \delta _{{\rm{max}}}} = \linebreak = \mathop {{\rm{sup}}}\limits_t \left\| {\dot \delta \left( t \right)} \right\|$ such that}
\begin{displaymath}
\left\| {\delta \left( t \right)} \right\| \le {\delta _{{\rm{max}}}} < \infty {\rm{, }}\left\| {\dot \delta \left( t \right)} \right\| \le {\dot \delta _{{\rm{max}}}} < \infty .    
\end{displaymath}

\emph{\textbf{Assumption 2.} There exist matrices $L \in {\mathbb{R}^{n \times p}}{\rm{,\;}}Q \in {\mathbb{R}^{n \times n}}{\rm{,\;}}P \in {\mathbb{R}^{n \times n}}{\rm{,\;}}M \in {\mathbb{R}^{s \times p}}$, which satisfy the following set of equations:}
\begin{equation}\label{eq2}
\begin{array}{l}
{\left( {A + LC} \right)^{\top}}P + P\left( {A + LC} \right) =  - Q{\rm{,}}\\
{D^{\top}}P = MC.
\end{array}    
\end{equation}

\emph{\textbf{Assumption 3.} For all $\theta  \in {D_\theta }$ any $u\left( t \right) \in {D_u}$ ensures}
\begin{displaymath}
\begin{array}{l}
\left\| {G\left( {y{\rm{,\;}}u} \right)} \right\| \le {G_{{\rm{max}}}} = \mathop {{\rm{sup}}}\limits_t \left\| {G\left( {y{\rm{,\;}}u} \right)} \right\| < \infty {\rm{,}}\\
\left\| {\dot G\left( {y{\rm{,\;}}u} \right)} \right\| \le {{\dot G}_{{\rm{max}}}} = \mathop {{\rm{sup}}}\limits_t \left\| {\dot G\left( {y{\rm{,\;}}u} \right)} \right\| < \infty .
\end{array}    
\end{displaymath}

The aim is to obtain the estimates of state $x\left( t \right)$, perturbation $\delta \left( t \right)$ and parameters $\theta $ so that the following inequalities hold:
\renewcommand{\theequation}{3a}
\begin{equation}\label{eq3a}
\mathop {{\rm{lim}}}\limits_{t \to \infty } \left\| {\tilde x\left( t \right)} \right\| \le {\varepsilon _x}{\rm{,\;}}\mathop {{\rm{lim}}}\limits_{t \to \infty } \left\| {\tilde \delta \left( t \right)} \right\| \le {\varepsilon _\delta }{\rm{,}}
\end{equation}
\renewcommand{\theequation}{3b}
\begin{equation}\label{eq3b}
\mathop {{\rm{lim}}}\limits_{t \to \infty } \left\| {\tilde \theta \left( t \right)} \right\| \le {\varepsilon _\theta }{\rm{,}}
\end{equation}
\renewcommand{\theequation}{\arabic{equation}}
where $\tilde x\left( t \right) = \hat x\left( t \right) - x\left( t \right)$ is the state reconstruction error, $\tilde \delta \left( t \right) = \hat \delta \left( t \right) - \delta \left( t \right)$ denotes the disturbance reconstruction error, $\tilde \theta \left( t \right) = \hat \theta \left( t \right) - \theta $ stands for the parametric error, \linebreak ${\varepsilon _x} > 0{\rm{,\;}}{\varepsilon _\delta } > 0{\rm{,\;}}{\varepsilon _\theta } > 0$ are arbitrarily small scalars.
\setcounter{equation}{3}

Assumption 1 is conventional for the adaptive observers design problems. In accordance with the results of \cite{b23}, if the condition $\phi \left( {y{\rm{,\;}}u} \right) + G\left( {y{\rm{,\;}}u} \right)\theta  \equiv 0$ is met, then assumption 2 allows one to ensure that \eqref{eq3a} hold for any ${\varepsilon _x} > 0{\rm{,\;}}{\varepsilon _\delta } > 0$. Assumption 3 is required to design the proposed estimation law for $\theta $. Unlike [23], the system \eqref{eq1} includes a nonlinearity $\phi \left( {y{\rm{,\;}}u} \right)$ and a parametric uncertainty $G\left( {y{\rm{,\;}}u} \right)\theta $. Contrary to \cite{b19,b20,b21}, the aim of this study is to guarantee the convergence of the parametric error $\tilde \theta \left( t \right)$ to arbitrarily small neighborhood of zero. In contrast to [17, 18], the internal model for $\delta \left( t \right)$ is unknown.

\section{Main result}
The state and disturbance of the system \eqref{eq1} is proposed to be reconstructed with the help of the following adaptive observer:
\begin{equation}\label{eq4}
\begin{array}{l}
\hat \delta \left( t \right) =  - \mu M\left( {\hat y\left( t \right) - y\left( t \right)} \right){\rm{,}}\\
\dot {\hat x}\left( t \right) \!=\! A\hat x\left( t \right) + \phi \left( {y{\rm{,\;}}u} \right) + G\left( {y{\rm{,\;}}u} \right)\hat \theta \left( t \right) + D\hat \delta \left( t \right){\rm{ + }}\\
+L\left( {\hat y\left( t \right) - y\left( t \right)} \right){\rm{,}}\\
\hat y\left( t \right) = C\hat x\left( t \right){\rm{,\;}}\hat x\left( {{t_0}} \right) = {{\hat x}_0}{\rm{,}}
\end{array}    
\end{equation}
where $\mu  > 0$ is a sufficiently large scalar, and $L$ and $M$ satisfy the system \eqref{eq2}.

To achieve the goal \eqref{eq3a}, \eqref{eq3b} via observer \eqref{eq4}, the unknown parameters $\theta $ are required to be identified in such a way that the parametric error $\tilde \theta \left( t \right)$ converges asymptotically to an arbitrarily small neighborhood of zero. Subsection A describes the estimation law that provides this property, and in Subection B the convergence of the errors $\tilde x\left( t \right)$ and $\tilde \delta \left( t \right)$ is analyzed when \eqref{eq4} and the proposed estimation law are used.

\subsection{Robust estimation of unknown parameters}
It is proposed to solve the problem of the parameter identification by means of the estimation law \cite{b22}, which ensures stated goal \eqref{eq3b} achievement in case of the external perturbation. To apply such a law, we first obtain a linear regression equation that shows the relation between the measured signals and the unknown parameters.

To do so, the following dynamic filters are introduced:
\begin{equation}\label{eq5}
\begin{array}{l}
\dot \chi \left( t \right) = {A_K}\chi \left( t \right) + Ky\left( t \right){\rm{,\;}}\chi \left( {{t_0}} \right) = {\chi _0}{\rm{,}}\\
\dot P\left( t \right) = {A_K}P\left( t \right) + \phi \left( {y{\rm{,\;}}u} \right){\rm{,\;}}P\left( {{t_0}} \right) = {0_n}{\rm{,}}\\
\dot \Omega \left( t \right) = {A_K}\Omega \left( t \right) + G\left( {y{\rm{,\;}}u} \right){\rm{,\;}}\Omega \left( {{t_0}} \right) = {0_{n \times q}}{\rm{,}}\\
{{\dot \Phi }_K}\left( t \right) = {A_K}{\Phi _K}\left( t \right){\rm{,\;}}{\Phi _K}\left( {{t_0}} \right) = {I_{n \times n}}{\rm{,}}
\end{array}    
\end{equation}
where $K \in {\mathbb{R}^{n \times p}}$ makes the matrix ${A_K} = A - KC$ be a Hurwitz one.

Based on the filters \eqref{eq5} state, the following error is considered:
\begin{equation}\label{eq6}
e\left( t \right) = \chi \left( t \right) + P\left( t \right) + \Omega \left( t \right)\theta  - x\left( t \right).    
\end{equation}

Owing to \eqref{eq1} and \eqref{eq5}, the derivative of \eqref{eq6} is written:
\begin{equation}\label{eq7}
\begin{array}{l}
\dot e\left( t \right) = {A_K}\chi \left( t \right) + Ky\left( t \right) + {A_K}P\left( t \right) + \phi \left( {y{\rm{,\;}}u} \right) + \\
 + \left[ {{A_K}\Omega \left( t \right) + G\left( {y{\rm{,\;}}u} \right)} \right]\theta  - Ax\left( t \right) - \phi \left( {y{\rm{,\;}}u} \right) -\\ - G\left( {y{\rm{,\;}}u} \right)\theta  - D\delta \left( t \right) = \\
 = {A_K}\left( {\chi \left( t \right) + P\left( t \right) + \Omega \left( t \right)\theta  - x\left( t \right)} \right) - D\delta \left( t \right) = \\
 = {A_K}e\left( t \right) - D\delta \left( t \right){\rm{,\;}}e\left( {{t_0}} \right) = {\chi _0} - {x_0}.
\end{array}    
\end{equation}

The solution of the differential equation \eqref{eq7} is obtained as:
\begin{equation}\label{eq8}
e\left( t \right) = {\Phi _K}\left( t \right){e_0} + {\delta _f}\left( t \right){\rm{,}}    
\end{equation}
where
\begin{displaymath}
{\dot \delta _f}\left( t \right) = {A_K}{\delta _f}\left( t \right) - D\delta \left( t \right){\rm{,\;}}{\delta _f}\left( {{t_0}} \right) = {0_n}.    
\end{displaymath}

Having substituted \eqref{eq8} into \eqref{eq6} and multiplied the result by ${\cal L}C$ (where ${\cal L} = {\begin{bmatrix}
1&1& \cdots &{{1_p}}
\end{bmatrix}}$), it is written:
\begin{equation}\label{eq9}
\begin{array}{c}
{\cal L}C\left[ {{\Phi _K}\left( t \right){e_0} + {\delta _f}\left( t \right)} \right] = {\cal L}C\chi \left( t \right) + \\
+ {\cal L}CP\left( t \right) + {\cal L}C\Omega \left( t \right)\theta  - {\cal L}Cx\left( t \right) \Rightarrow \\
z\left( t \right) = {\varphi ^{\top}}\left( t \right)\theta  + w\left( t \right)
\end{array}    
\end{equation}
where
\begin{displaymath}
\begin{gathered}
  z\left( t \right) = \mathcal{L}\left[ {y\left( t \right) - C\chi \left( t \right) - CP\left( t \right)} \right]{\text{, }} \\ 
  {\varphi ^{\top}}\left( t \right) = \mathcal{L}C\Omega \left( t \right){\text{, }}w\left( t \right) =  - \mathcal{L}C{\Phi _K}\left( t \right){e_0} - \mathcal{L}C{\delta _f}\left( t \right). \\ 
\end{gathered}
\end{displaymath}

To identify the parameters of the regression equation \eqref{eq9}, we apply the estimation law from \cite{b22}, which guarantees that the following holds:
\begin{equation}\label{eq10}
\mathop {{\rm{lim}}}\limits_{t \to \infty } \left\| {\tilde \theta \left( t \right)} \right\| \le \varepsilon \left( T \right){\rm{,\;}}\mathop {{\rm{lim}}}\limits_{T \to \infty } \varepsilon \left( T \right) = 0{\rm{,}}
\end{equation}
where $T > 0$ is some parameter of the identification algorithm, and $\varepsilon {\rm{:\;}}{\mathbb{R}_ + } \mapsto {\mathbb{R}_ + }$.

{To introduce} the above-mentioned law in accordance with \cite{b22}, a linear dynamic filter ${\cal H}\left( s \right) = {\textstyle{\alpha  \over {s + \alpha }}}{\rm{,\;}}\alpha  > 0$ is applied to the left- and right-hand sides of equation \eqref{eq1}:
\begin{equation}\label{eq11}
{z_f}\left( t \right) = \varphi _f^{\top}\left( t \right)\theta  + {w_f}\left( t \right){\text{,}}
\end{equation}
where
\begin{displaymath}
\begin{gathered}
        {z_f}\left( t \right){\text{:}} = \mathcal{H}\left( s \right)z\left( t \right){\text{, }}{\varphi _f}\left( t \right){\text{:}} = \mathcal{H}\left( s \right)\varphi \left( t \right){\text{,}}\;\\{w_f}\left( t \right){\text{:}} = \mathcal{H}\left( s \right)w\left( t \right).
\end{gathered}
\end{displaymath}

Subtracting \eqref{eq11} from \eqref{eq9}, it is obtained:
\begin{equation}\label{eq12}
\tilde{z}\left( t \right) = {\phi ^{\top}}\left( t \right)\Theta  + f\left( t \right){\rm{,}}    
\end{equation}
where
\begin{displaymath}
\begin{array}{c}
\tilde{z}\left(t\right)=z\left( t \right) - {z_f}\left( t \right){\rm{,}}\\
\phi \left( t \right){\rm{:}} = { {\begin{bmatrix}
{{\varphi ^{\top}}\left( t \right)}&{\varphi _f^{\top}\left( t \right)}
\end{bmatrix}}^{\top}}{\rm{, }}\\
f\left( t \right){\rm{:}} = w\left( t \right) - {w_f}\left( t \right){\rm{,\;}}\Theta {\rm{:}} = {\begin{bmatrix}
\theta \\
{ - \theta }
\end{bmatrix}}.
\end{array}    
\end{displaymath}

{Afterwards,} the regression equation \eqref{eq12} is extended as follows:
\begin{equation}\label{eq13}
\begin{array}{l}
\dot Y\left( t \right) = {\textstyle{1 \over T}}\left[ \phi \left( t \right){\tilde{z}\left( t \right)} - \phi \left( {t - T} \right) {\tilde{z}\left( {t - T} \right)} \right]{\rm{,\;}}\\
\dot \Phi \left( t \right) = {\textstyle{1 \over T}}\left[ {\phi \left( t \right){\phi ^{\top}}\left( t \right) - \phi \left( {t - T} \right){\phi ^{\top}}\left( {t - T} \right)} \right]{\rm{,\;}}\\
Y\left( {{t_0}} \right) = {0_{2q}}{\rm{,\;}}\Phi \left( {{t_0}} \right) = {0_{2q \times 2q}}.
\end{array}    
\end{equation}

{Considering \eqref{eq12}, it is obvious that the signals \eqref{eq13} satisfy the following equation:}
\begin{equation}\label{eq14}
Y\left( t \right) = \Phi \left( t \right)\Theta  + W\left( t \right){\rm{,}}
\end{equation}
where
\begin{displaymath}
\begin{gathered}
\dot W\left( t \right) = {\textstyle{1 \over T}}\left[ {\phi \left( t \right)f\left( t \right) - \phi \left( {t - T} \right)f\left( {t - T} \right)} \right]{\rm{,\;}}
W\left( {{t_0}} \right) = {0_{2q}}.        
\end{gathered}
\end{displaymath}

Disturbance term in \eqref{eq14} always admits the decomposition:
\begin{equation}\label{eq15}
W\left( t \right){\rm{:}} = {{\cal L}_1}{\cal L}_1^{\top}W\left( t \right){\rm{ + }}{{\cal L}_2}{\cal L}_2^{\top}W\left( t \right){\rm{,}}    
\end{equation}
where ${{\cal L}_1} \in {\mathbb{R}^{2q \times 2m}}$ and ${{\cal L}_2} \in {\mathbb{R}^{2q \times \left( {2q - 2m} \right)}}$ meet the following conditions:
\begin{displaymath}
\begin{array}{c}
{\cal L}_1^{\top}{{\cal L}_1} = {I_{2m \times 2m}}{\rm{,\;}}{\cal L}_2^{\top}{{\cal L}_2} = {I_{\left( {2q - 2m} \right) \times \left( {2q - 2m} \right)}}{\rm{, }}\\
{{\cal L}_1}{\cal L}_1^{\top} + {{\cal L}_2}{\cal L}_2^{\top} = {I_{2q \times 2q}}{\rm{,}}
\end{array}
\end{displaymath}
and $2\left( {q - m} \right)$ is the number of elements of the vector $\phi \left( t \right)$, for which the \emph{independence} condition holds:
\begin{equation}\label{eq16}
\mathop {{\rm{lim}}}\limits_{T \to \infty } {\textstyle{1 \over T}}\int\limits_{{\rm{max}}\left\{ {{t_0}{\rm{,\;}}t - T} \right\}}^t {{\phi _i}\left( s \right)f\left( s \right)ds}  = 0.
\end{equation}

Multiplication of equation \eqref{eq14} by ${\rm{adj}}\left\{ {\Phi \left( t \right)} \right\}$ yields:
\begin{equation}\label{eq17}
{\cal Y}\left( t \right) = \Delta \left( t \right)\Theta  + {{\cal W}_1}\left( t \right) + {{\cal W}_2}\left( t \right){\rm{,}}
\end{equation}
where
\begin{displaymath}
\begin{array}{c}
{\cal Y}\left( t \right){\rm{:}} = {\rm{adj}}\left\{ {\Phi \left( t \right)} \right\}Y\left( t \right){\rm{,\;}}\Delta \left( t \right){\rm{:}} = {\rm{det}}\left\{ {\Phi \left( t \right)} \right\}{\rm{,\;}}\\
{{\cal W}_1}\left( t \right){\rm{:}} = {\rm{adj}}\left\{ {\Phi \left( t \right)} \right\}{{\cal L}_1}{\cal L}_1^{\top}W\left( t \right){\rm{,\;}}\\
{{\cal W}_2}\left( t \right){\rm{:}} = {\rm{adj}}\left\{ {\Phi \left( t \right)} \right\}{{\cal L}_2}{\cal L}_2^{\top}W\left( t \right).
\end{array}
\end{displaymath}

Owing to the definition of $\Theta $, for any $q \ge m \ge 1$ there exists and is known an \emph{annihilator} ${{\cal H}^{\top}} \in {\mathbb{R}^{2m \times 2q}}$ of full column rank such that
\begin{equation}\label{eq18}
{{\cal H}^{\top}}\Theta  = 0.    
\end{equation}

Considering \eqref{eq18}, the multiplication of \eqref{eq17} firstly by $\mathcal{H}^{\top}$ and then by ${\text{adj}}\left\{ {{\mathcal{H}^{\top}}{\text{adj}}\left\{ {\Phi \left( t \right)} \right\}{\mathcal{L}_1}} \right\}$ yields:
\begin{equation}\label{eq19}
\begin{gathered}
    \mathcal{N}\left( t \right) = \mathcal{M}\left( t \right)\mathcal{L}_1^{\top}W\left( t \right) + \hfill\\\hfill
    + {\text{adj}}\left\{ {{\mathcal{H}^{\top}}{\text{adj}}\left\{ {\Phi \left( t \right)} \right\}{\mathcal{L}_1}} \right\}{\mathcal{H}^{\top}}{\mathcal{W}_2}\left( t \right){\text{,}}
\end{gathered}    
\end{equation}	
where
\begin{displaymath}
\begin{gathered}
  \mathcal{N}\left( t \right){\text{:}} = {\text{adj}}\left\{ {{\mathcal{H}^{\top}}{\text{adj}}\left\{ {\Phi \left( t \right)} \right\}{\mathcal{L}_1}} \right\}{\mathcal{H}^{\top}}\mathcal{Y}\left( t \right){\text{, }} \hfill \\
  \mathcal{M}\left( t \right){\text{:}} = {\text{det}}\left\{ {{\mathcal{H}^{\top}}{\text{adj}}\left\{ {\Phi \left( t \right)} \right\}{\mathcal{L}_1}} \right\}. \hfill \\ 
\end{gathered}
\end{displaymath}
 
Now we are in position to \emph{annihilate} the part of perturbation term in \eqref{eq14} via simple substitution. For that purpose, equation \eqref{eq14} is multiplied by $\mathcal{M}\left( t \right)$, and ${\mathcal{L}_1}\mathcal{N}\left( t \right)$ is subtracted from the obtained result to write:
\begin{equation}\label{eq20}
\begin{gathered}
    \lambda \left( t \right) = \Omega \left( t \right)\Theta  +\hfill\\+ \left[ \mathcal{M}\left( t \right){\mathcal{L}_2} - {\mathcal{L}_1}
   {\text{adj}}\left\{ {{\mathcal{H}^{\top}}{\text{adj}}\left\{ {\Phi \left( t \right)} \right\}{\mathcal{L}_1}} \right\}\times\right.\hfill\\
    \left. \times{\mathcal{H}^{\top}}{\text{adj}}\left\{ {\Phi \left( t \right)} \right\}{\mathcal{L}_2} \right]\mathcal{L}_2^{\top}W\left( t \right){\text{,}}\hfill
\end{gathered}
\end{equation}
where
\begin{displaymath}
\begin{gathered}
  \lambda \left( t \right){\text{:}} = \mathcal{M}\left( t \right)Y\left( t \right) - {\mathcal{L}_1}\mathcal{N}\left( t \right){\text{,}} \hfill \\
  \Omega \left( t \right){\text{:}} = \mathcal{M}\left( t \right)\Phi \left( t \right). \hfill \\ 
\end{gathered}
\end{displaymath}
 
To obtain the regression equation with a regressor, which derivative is directly measurable, we use the following simple filtration ($k > 0$):
\begin{equation}\label{eq21}
\begin{array}{l}
{{\dot \Omega }_f}\left( t \right) =  - k{\Omega _f}\left( t \right) + k\Omega \left( t \right){\rm{,\;}}{\Omega _f}\left( {{t_0}} \right) = {0_{2q \times 2q}}{\rm{,}}\\
{{\dot \lambda }_f}\left( t \right) =  - k{\lambda _f}\left( t \right) + k\lambda \left( t \right){\rm{,\;}}{\lambda _f}\left( {{t_0}} \right) = {0_{2q}}{\rm{,}}
\end{array}    
\end{equation}
then, to convert \eqref{eq20} into a set of separate scalar regression equations, the signal ${\lambda _f}\left( t \right)$ is multiplied by ${\rm{adj}}\left\{ {{\Omega _f}} \right\}$:
\begin{equation}\label{eq22}
\begin{gathered}
\Lambda \left( t \right) = \omega \left( t \right)\Theta  + d\left(t\right){\rm{,}}
\end{gathered}
\end{equation}
where
\begin{displaymath}
\begin{array}{l}
\Lambda \left( t \right){\rm{:}} = {\rm{adj}}\left\{ {{\Omega _f}} \right\}{\lambda _f}\left( t \right){\rm{,}}\\
\omega \left( t \right){\rm{:}} = {\rm{det}}\left\{ {{\Omega _f}} \right\},\\
d\left(t\right)={\text{adj}}\left\{ {{\Omega _f}} \right\}\tfrac{k}{{s + k}}\left\{ \left[ \mathcal{M}\left( t \right){\mathcal{L}_2} - {\mathcal{L}_1}\times\right.\right.\\
\left.\left.\times{\text{adj}}\left\{ {{\mathcal{H}^{\top}}{\text{adj}}\left\{ {\Phi \left( t \right)} \right\}{\mathcal{L}_1}} \right\}{\mathcal{H}^{\top}}{\text{adj}}\left\{ {\Phi \left( t \right)} \right\}{\mathcal{L}_2} \right]\mathcal{L}_2^{\top}W\left( t \right) \right\}{\text{,}}    
\end{array}    
\end{displaymath}

Using \eqref{eq22}, the following estimation law is introduced:
\begin{equation}\label{eq23}
\begin{array}{l}
\hat \theta \left( t \right) = \hat \kappa \left( t \right){{\cal L}_0}\Lambda \left( t \right){\rm{,}}\\
\dot {\hat \kappa} \left( t \right) =  - \gamma \omega \left( t \right)\left( {\omega \left( t \right)\hat \kappa \left( t \right) - 1} \right) - \dot \omega \left( t \right){{\hat \kappa }^2}\left( t \right){\rm{,}}\;\\
\dot \omega \left( t \right) = {\rm{tr}}\left( {{\rm{adj}}\left\{ {{\Omega _f}\left( t \right)} \right\}{{\dot \Omega }_f}\left( t \right)} \right){\rm{, }}\\
\hat \kappa \left( {{t_0}} \right) = {{\hat \kappa }_0}{\rm{,\;}}\omega \left( {{t_0}} \right) = 0{\rm{,}}
\end{array}    
\end{equation}
where ${{\cal L}_0} = {\begin{bmatrix}
{{I_{q \times q}}}&{{0_{q \times q}}}
\end{bmatrix}}$, $\gamma  > 0$, {and $\hat\kappa\left(t\right)$ is the estimate of $\omega^{-1}\left(t\right)$}.
 
 The conditions, under which the goal \eqref{eq10} is achieved when the estimation law \eqref{eq23} is applied, are given in the following theorem.

\textbf{Theorem 1.} \emph{Suppose that $\varphi \left( t \right){\rm{,\;}}w\left( t \right)$ are bounded and assume that:}
\begin{enumerate}
    \item[\textbf{C1)}] \emph{there exist (possibly do not unique) $T \ge {T_f} > 0$ and $\underline \alpha   \ge \overline \alpha   > 0$ such that for all $t \ge {T_f}$ it holds that}
    \begin{equation}\label{eq24}
        \hspace{-20pt}0 < \underline \alpha  {I_{2q}} \le {\textstyle{1 \over T}}\int\limits_{{\rm{max}}\left\{ {{t_0}{\rm{,\;}}t - T} \right\}}^t {\phi \left( s \right){\phi ^{\top}}\left( s \right)ds}  \le \overline \alpha  {I_{2q}}{\rm{,}}
    \end{equation}
    \item[\textbf{C2)}] \emph{the equality \eqref{eq16} holds for $2\left( {q - m} \right) \ge 2$ elements of $\phi \left( t \right)$,}
    \item[\textbf{C3)}] \emph{eliminators ${{\cal L}_1} \in {\mathbb{R}^{2q \times 2m}}{\rm{,\;}}{{\cal L}_2} \in {\mathbb{R}^{2q \times \left( {2q - 2m} \right)}}$ are exactly known and such that there exist (possibly do not unique) $T \ge {T_f} > 0$ and $\underline \beta   \ge \overline \beta   > 0$ such that for all $t \ge {T_f}$ it holds that:}
    \begin{equation*}
        \hspace{-20pt}0 \!<\! \underline \beta\! \le \!\left| {{\rm{det}}\!\left\{\! {{{\cal H}^{\top}}{\rm{adj}}\left\{ {{\textstyle{1 \over T}}\!\!\!\!\int\limits_{{\rm{max}}\left\{ {{t_0}{\rm{,\;}}t - T} \right\}}^t {\!\!\!\!\!\!\!\!\!\!\phi \left( s \right){\phi ^{\top}}\left( s \right)ds} } \right\}\!{{\cal L}_1}} \!\right\}} \right| \!\le\! \overline \beta  {\rm{,}}
    \end{equation*}
    \item[\textbf{C4)}] \emph{$\gamma  > 0$ is chosen so that there exists $\eta  > 0$ such that}
    \begin{displaymath}
        \gamma {\omega ^3}\left( t \right) + \omega \left( t \right)\dot \omega \left( t \right)\hat \kappa \left( t \right) + \dot \omega \left( t \right) \ge \eta \omega \left( t \right) > 0\;\forall t \ge {T_f}.
    \end{displaymath}
\end{enumerate}

\emph{Then the estimation law (23) ensures the existence of $\varepsilon {\rm{:\;}}{\mathbb{R}_ + } \mapsto {\mathbb{R}_ + }$ such that the limits (10) hold.}

\emph{Proof of theorem 1 is given in} [22].

~

Therefore, if the conditions \eqref{eq10} are met, then we immediately conclude that there exists an arbitrarily small scalar ${\varepsilon _\theta } > 0$ that meets inequality \eqref{eq3b}. Requirement \textbf{C1} is a sufficient identifiability condition for the parameters $\Theta $ in the perturbation-free case $W\left( t \right) = 0$. Requirements \textbf{C2} and \textbf{C3} describe the identifiability conditions for the value of $\mathop {{\rm{lim}}}\limits_{T \to \infty } {\cal L}_1^{\top}W\left( t \right)$, and \textbf{C4} is necessary to ensure $\hat \kappa \left( t \right) \to \linebreak \to {\omega ^{ - 1}}\left( t \right)$ as $t \to \infty $. Condition \eqref{eq16} is met if the regressor ${\phi _i}\left( t \right)$ spectrum does not have common frequencies with the perturbation $f\left( t \right)$ one, and therefore, if the identifiability conditions \textbf{C1} and \textbf{C2} are satisfied, then the estimation law \eqref{eq23} guarantees that \eqref{eq10} is met so long as the spectrum of at least one element of the regressor $\varphi\left(t\right)$ has no common frequencies with the perturbation. Note that, if \textbf{C1}-\textbf{C4} are satisfied, the parametric convergence \eqref{eq10} is achieved regardless of the output matching condition \eqref{eq2}.

\subsection{Stability analysis}
The properties of the observer \eqref{eq4} enhanced by the estimation law \eqref{eq23} are given in the following theorem.

\textbf{Theorem 2.} \emph{Let assumptions 1-3 and conditions \textbf{C1}-\textbf{C4} be met. Then for all ${\varepsilon _x} > 0$ and ${\varepsilon _\delta } > 0$ there exists ${\mu ^*} > 0$ such that for all $\mu  \ge {\mu ^*}$ inequality \eqref{eq3a} holds.}

\emph{Proof.} The differential equation for $\tilde x\left( t \right)$ is written as:
\begin{equation}\label{eq26}
\begin{array}{l}
\dot {\tilde x} = A\hat x \!+\! \phi \left( {y{\rm{,\;}}u} \right) \!+\! G\left( {y{\rm{,\;}}u} \right)\hat \theta 
\!+\! D\hat \delta {\rm{ + }}L\left( {\hat{y} - y} \right) \!-\!\\- Ax- \phi \left( {y{\rm{,\;}}u} \right)
- G\left( {y{\rm{,\;}}u} \right)\theta  - D\delta  = \\=\left( {A + LC} \right)\tilde x
+ G\left( {y{\rm{,\;}}u} \right)\tilde \theta  \!+\! D\tilde \delta  = \\ \!=\! \left( {A + LC - \mu D{D^{\top}}P} \right)\tilde x + G\left( {y{\rm{,\;}}u} \right)\tilde \theta  - D\delta .
\end{array}
\end{equation}

The following quadratic form is introduced:
\begin{equation}\label{eq27}
V = {\tilde x^{\top}}P\tilde x.    
\end{equation}

Owing to \eqref{eq26}, the derivative of \eqref{eq27} is written as:
\begin{equation}\label{eq28}
\begin{array}{l}
\dot V = {{\tilde x}^{\top}}\left[ {{{\left( {A + LC} \right)}^{\top}}P + P\left( {A + LC} \right)} \right]\tilde x -\\
-\! 2\mu {{\tilde x}^{\top}}PD{D^{\top}}P\tilde x \!-\! 2{{\tilde x}^{\top}}PD\delta  + 2{{\tilde x}^{\top}}PG\left( {y{\rm{,\;}}u} \right)\tilde \theta  = \\
 =  - {{\tilde x}^{\top}}Q\tilde x - 2\mu {{\tilde x}^{\top}}PD{D^{\top}}P\tilde x -\\\hfill
 - 2{{\tilde x}^{\top}}PD\delta  + 2{{\tilde x}^{\top}}PG\left( {y{\rm{,\;}}u} \right)\tilde \theta .
\end{array}    
\end{equation}

As the below-given inequalities hold:
\begin{equation}\label{eq29}
\begin{array}{l}
2\left\| {{D^{\top}}P\tilde x} \right\|{\delta _{{\rm{max}}}} \le \gamma _1^{ - 1}{\left\| {{D^{\top}}P\tilde x} \right\|^2}\delta _{{\rm{max}}}^2 + {\gamma _1}{\rm{,}}\\
2\left\| {{{\tilde x}^{\top}}PG\left( {y{\rm{,\;}}u} \right)\tilde \theta } \right\| \le \left\| {\tilde x} \right\| \cdot \left\| P \right\|{G_{{\rm{max}}}} \cdot \left\| {\tilde \theta } \right\| \le \\
\le \gamma _2^{ - 1}{\left\| {\tilde x} \right\|^2}{\left\| {\tilde \theta } \right\|^2} + {\gamma _2}{\left\| P \right\|^2}G_{{\rm{max}}}^2{\rm{,}}
\end{array}    
\end{equation}
then the following upper bound for \eqref{eq28} is obtained:
\begin{equation}\label{eq30}
\begin{gathered}
\dot V \le  - \left[ {\lambda _{{\rm{min}}}}\left( Q \right) - \gamma _2^{ - 1}{{\left\| {\tilde \theta } \right\|}^2} \right]{\left\| {\tilde x} \right\|^2} -\hfill\\
 - \left[ {2\mu  - \gamma _1^{ - 1}\delta _{{\rm{max}}}^2} \right]{\left\| {{D^{\top}}P\tilde x} \right\|^2} + {\gamma _1} + {\gamma _2}{\left\| P \right\|^2}G_{{\rm{max}}}^2{\rm{,}}        
\end{gathered}
\end{equation}
where we use the estimates from Assumption 3.

As, using the results of theorem 1, we have $\mathop {{\rm{lim}}}\limits_{T \to \infty } \mathop {{\rm{lim}}}\limits_{t \to \infty } \left\| {\tilde \theta \left( t \right)} \right\| = 0$, then, owing to continuity of the error $\tilde \theta \left( t \right)$ with respect to $t$ and $T$, there exists a time instant ${t_{\tilde x}} \ge {t_0}$ and a sufficiently large scalar $T > 0$ such that for all $t \ge {t_{\tilde x}}$ it holds that
\begin{equation}\label{eq31}
{\lambda _{{\rm{min}}}}\left( Q \right) - \gamma _2^{ - 1}{\left\| {\tilde \theta } \right\|^2} \ge {c_{\tilde x}} > 0{\rm{,}}    
\end{equation}
and, therefore, choosing
\begin{equation}\label{eq32}
\mu  \ge {\mu ^ * } = {\textstyle{1 \over 2}}\gamma _1^{ - 1}\delta _{{\rm{max}}}^2
\end{equation}
the upper bound for the derivative \eqref{eq30} is rewritten as follows:
\begin{equation}\label{eq33}
\dot V \le  - {\textstyle{{{c_{_{_{\tilde x}}}}} \over {{\lambda _{{\rm{max}}}}\left( P \right)}}}V + {\gamma _1} + {\gamma _2}{\left\| P \right\|^2}G_{{\rm{max}}}^2.
\end{equation}

The solution of the differential equation \eqref{eq33} is obtained as:
\begin{equation}\label{eq34}
\begin{gathered}
\left\| {\tilde x\left( t \right)} \right\| \le \sqrt {{\textstyle{{{\lambda _{{\rm{max}}}}\left( P \right)} \over {{\lambda _{{\rm{min}}}}\left( P \right)}}}} {e^{{\textstyle{{ - {c_{\tilde x}}} \over {2{\lambda _{{\rm{max}}}}\left( P \right)}}}\left( {t - {t_{\tilde x}}} \right)}}\left\| {\tilde x\left( {{t_{\tilde x}}} \right)} \right\| +\hfill\\\hfill
+ \sqrt {{\textstyle{{\left[ {{\gamma _1} + {\gamma _2}{{\left\| P \right\|}^2}G_{{\rm{max}}}^2} \right]{\lambda _{{\rm{max}}}}\left( P \right)} \over {{c_{_{\tilde x}}}{\lambda _{{\rm{min}}}}\left( P \right)}}}}.    
\end{gathered}
\end{equation}

The signal $\sigma \left( t \right) = \dot {\tilde x}\left( t \right)$ is introduced, as well as the differential equation for it:
\begin{equation}\label{eq35}
\begin{array}{l}
\dot \sigma = \left( {A + LC - \mu D{D^{\top}}P} \right)\sigma - D\dot \delta \hfill+\\\hfill+ \dot G\left( {y{\rm{,\;}}u} \right)\tilde \theta + G\left( {y{\rm{,\;}}u} \right)\dot {\tilde \theta}.
\end{array}
\end{equation}

Owing to \eqref{eq35}, the derivative of ${V_\sigma } = {\sigma ^{\top}}P\sigma $ is obtained as:
\begin{equation}\label{eq36}
\begin{array}{c}
{{\dot V}_\sigma } =  - {\sigma ^{\top}}Q\sigma  - 2\mu {\sigma ^{\top}}PD{D^{\top}}P\sigma  - 2{\sigma ^{\top}}PD\dot \delta  + \hfill\\\hfill
 + 2{\sigma ^{\top}}P\dot G\left( {y{\rm{,\;}}u} \right)\tilde \theta  + 2{\sigma ^{\top}}PG\left( {y{\rm{,\;}}u} \right)\dot {\tilde \theta} .
\end{array}
\end{equation}

If the conditions \textbf{C1}-\textbf{C3} hold, then for all $t \ge {T_f}$ we have:
\begin{equation*}
\begin{gathered}
    \dot {\tilde \theta} \left( t \right) = s\left\{ \Omega _f^{ - 1}\left( t \right)\tfrac{k}{{s + k}}\left\{ \left[ \mathcal{M}\left( t \right){\mathcal{L}_2} - {\mathcal{L}_1}\times\right.\right.\right.\hfill\\\hfill
    \left.\left.\left. \times {\text{adj}}\left\{ {{\mathcal{H}^{\top}}{\text{adj}}\left\{ {\Phi \left( t \right)} \right\}{\mathcal{L}_1}} \right\}{\mathcal{H}^{\top}}{\text{adj}}\left\{ {\Phi \left( t \right)} \right\}{\mathcal{L}_2} \right]\mathcal{L}_2^{\top}W\left( t \right) \right\} \right\}{\text{,}}
\end{gathered}
\end{equation*}
and, therefore, it holds that:
\begin{displaymath}
\mathop {{\rm{lim}}}\limits_{T \to \infty } \mathop {{\rm{lim}}}\limits_{t \to \infty } \left\| {\dot{\tilde {\theta}} \left( t \right)} \right\| = \mathop {{\rm{lim}}}\limits_{t \to \infty } \mathop {{\rm{lim}}}\limits_{T \to \infty } \left\| {\dot {\tilde \theta} \left( t \right)} \right\| = 0.    
\end{displaymath}

Equation \eqref{eq36} is similar to \eqref{eq28}, so, using assumptions 1 and 3, we repeat the derivation \eqref{eq29}-\eqref{eq34}, and, considering that:
\begin{enumerate}
    \item[\emph{i)}] the equalities 
    \begin{equation*}
        \mathop {{\rm{lim}}}\limits_{T \to \infty } \mathop {{\rm{lim}}}\limits_{t \to \infty } \left\| {\tilde \theta \left( t \right)} \right\| = 0,\;\mathop {{\rm{lim}}}\limits_{T \to \infty } \mathop {{\rm{lim}}}\limits_{t \to \infty } \left\| {\dot {\tilde \theta} \left( t \right)} \right\| = 0
    \end{equation*} hold according to results of theorem 1,
    \item[\emph{ii)}] $\tilde \theta \left( t \right)$ and $\dot {\tilde \theta} \left( t \right)$ are continuous with respect to $t$ and $T$,
\end{enumerate}
we obtain the following upper bound:
\begin{equation}\label{eq38}
\begin{array}{l}
\left\| {\sigma \left( t \right)} \right\| \le \sqrt {{\textstyle{{{\lambda _{{\rm{max}}}}\left( P \right)} \over {{\lambda _{{\rm{min}}}}\left( P \right)}}}} {e^{{\textstyle{{ - {c_\sigma }} \over {2{\lambda _{{\rm{max}}}}\left( P \right)}}}\left( {t - {t_\sigma }} \right)}} \left\| {\sigma \left( {{t_\sigma }} \right)} \right\|+\hfill\\\hfill+ \sqrt {{\textstyle{{{\gamma _3}{\lambda _{{\rm{max}}}}\left( P \right)} \over {{c_\sigma }{\lambda _{{\rm{min}}}}\left( P \right)}}}} {\rm{,}}
\end{array}
\end{equation}
where ${\gamma _3} > 0$ is an arbitrary small scalar and ${t_\sigma } \ge {t_0}$.

Equation for $\tilde \delta \left( t \right)$ is rewritten as:
\begin{equation}\label{eq39}
\begin{gathered}
\tilde \delta \left( t \right) \!=\! {D^{\rm{\dag }}}\left[ {\sigma \left( t \right) \!-\! \left( {A + LC} \right)\tilde x\left( t \right)\!-\! G\left( {y{\rm{,\;}}u} \right)\tilde \theta \left( t \right)} \right].
\end{gathered}
\end{equation}

As $\mathop {{\rm{lim}}}\limits_{T \to \infty } \mathop {{\rm{lim}}}\limits_{t \to \infty } \left\| {\tilde \theta \left( t \right)} \right\| = 0$, $\left\| {G\left( {y{\rm{,\;}}u} \right)} \right\| \le {G_{{\rm{max}}}}$, then the choice of the arbitrary parameters ${\gamma _1} > 0{\rm{,\;}}{\gamma _2} > 0{\rm{,}}$ ${\gamma _3} > 0$ for \eqref{eq34} and \eqref{eq38} allows one to conclude that \eqref{eq3a} holds for any ${\varepsilon _x} > 0$ and ${\varepsilon _\delta } > 0$, which was to be proved.

\section{Numerical experiments}
The proposed adaptive observer has been applied to reconstruct the state of the Duffing oscillator \cite{b20}:
\begin{equation}\label{eq40}
\begin{array}{l}
\dot x = {\begin{bmatrix}
0&1\\
1&{ - 0.2}
\end{bmatrix}}x +  {\begin{bmatrix}
0&0\\
u&{ - {y^3}}
\end{bmatrix}} {\begin{bmatrix}
{{\theta _1}}\\
{{\theta _2}}
\end{bmatrix}}  + {\begin{bmatrix}
1\\
0
\end{bmatrix}}\delta {\rm{,}}\\
y = {x_1}{\rm{,}}
\end{array}    
\end{equation}

The disturbance $\delta \left( t \right)$ and control $u\left( t \right)$ signals, as well as the parameters of the system \eqref{eq40}, observer \eqref{eq4} and filters \eqref{eq5}, \eqref{eq11}, \eqref{eq13}, \eqref{eq21} were picked as:
\begin{equation}\label{eq41}
\begin{array}{c}
\delta \left( t \right) \!=\! 0.5{\rm{sin}}\left( {2t} \right){\rm{,\;}}u\left( t \right) = A{\rm{cos}}\left( t \right){\rm{,\;}}\theta  \!=\! { {\begin{bmatrix}
1&3
\end{bmatrix}} ^{\top}}{\rm{,\;}}\\
{x_0} = { {\begin{bmatrix}
2&1
\end{bmatrix}} ^{\top}}{\rm{,\;}} K =  - L = {\begin{bmatrix}
{30.5749}\\
{64.3579}
\end{bmatrix}}{\rm{, }}\\
M = 28.644{\rm{,\;}}\mu=25{\rm{,\;}}\alpha  = 0.1{\rm{,\;}}T = 30{\rm{,\;}}k = 1{\rm{,}}
\end{array}
\end{equation}
where the magnitude of $A > 0$ will be defined below.

{The elimination ${{\cal L}_1}$ and annihilation ${{\cal H}^{\top}}$ matrices were chosen as follows:}
\begin{displaymath}
{\cal L}_1^{\top} = {\begin{bmatrix}
0&1&0&0\\
0&0&0&1
\end{bmatrix}},\;{{\cal H}^{\top}} = {\begin{bmatrix}
1&0&1&0\\
0&1&0&1
\end{bmatrix}}. 
\end{displaymath}

Following proof of theorem 1, meeting the conditions \textbf{C1} and \textbf{C3} immediately results in the existence of ${\overline T_f} \ge {T_f}$ and ${\omega _{{\rm{LB}}}} > 0$ such that $\left| {\omega \left( t \right)} \right| \ge {\omega _{{\rm{LB}}}} > 0$ for all $t \ge {\overline T_f}$. In its turn, such condition is required to achieve the convergence $\hat \kappa \left( t \right) \to {\omega ^{ - 1}}\left( t \right)$ as $t \to \infty $. Therefore, we first obtain the behavior of $\omega \left( t \right)$ for different values of $A$ of the control signal $u\left( t \right)$.

\begin{figure}[!thpb]
\begin{center}
\includegraphics[scale=0.8]{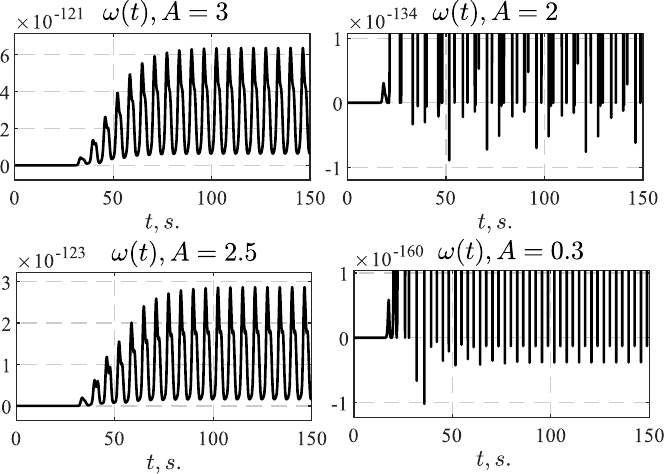}
\caption{{Dependence of $\omega \left( t \right)$ from $A$}} 
\end{center}
\end{figure}

Figure 1 shows that for fixed $T>0$ it is possible to meet the parametric convergence conditions \textbf{C1}, \textbf{C3} (see the proof of Theorem 1 in \cite{b22}) by choice of the input signal.

The adaptive gain of the estimation law \eqref{eq23} was set as follows:
$\gamma  = {10^{248}}{\rm{,}}$
and its large value is explained by the fact that $\omega \left( t \right) \in \left( {{{10}^{ - 125}}{\rm{,\;}}{{10}^{ - 120}}} \right)$ (see Fig.1).

Figure 2 presents behavior of $\tilde \theta \left( t \right)$ and $\left| {\tilde x\left( t \right)} \right|{\rm{,\;}}\left| {\tilde \delta \left( t \right)} \right|$ for $u\left( t \right) = 2.5{\rm{cos}}\left( {t} \right)$ and different values of the parameter $T  > 0$.
\begin{figure}[!thpb]
\begin{center}
\includegraphics[scale=0.55]{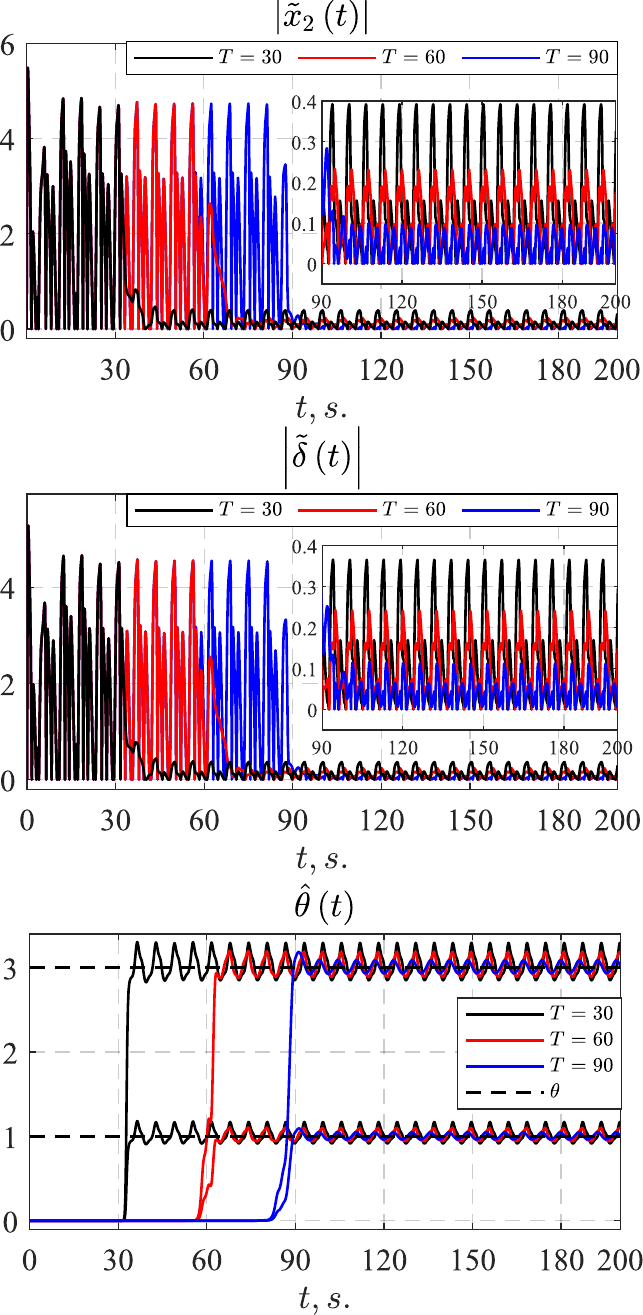}
\caption{{Behavior of errors $\left| {\tilde x_{2}\left( t \right)} \right|{\rm{,\;}}\left| {\tilde \delta \left( t \right)} \right|$ and $\tilde \theta \left( t \right)$ for different $T  > 0$}} 
\end{center}
\end{figure}
The obtained transients illustrate that appropriate choice of $T  > 0$ ensures that \eqref{eq3a} holds for any ${\varepsilon _\theta} > 0$. The simulation validates the conclusions of the theoretical analysis and shows that the conditions that are necessary to achieve \eqref{eq3a}, \eqref{eq3b} can be met by appropriate choice of a test signal (control).

\section{Conclusion}
A new adaptive observer is proposed for nonlinear systems with unknown input and parametric uncertainty. Unlike most existing solutions, the proposed approach provides: \linebreak 1) asymptotic identification of the unknown parameters with arbitrary accuracy in case the system is affected by an unknown but bounded perturbation, 2) convergence of state and perturbation estimates to an arbitrarily small neighborhood of the equilibrium point.

The application of the observer is possible if the system is stable (assumption 3), has a strictly positive real transfer function from perturbation to output (assumption 2), the perturbation is bounded together with its first derivative (assumption 1), and the parametric convergence conditions \textbf{C1}-\textbf{C4} are met. Inequalities \textbf{C1} and \textbf{C3} are similar to the condition of the regressor persistent excitation and are necessary to provide identifiability of the unknown parameters and perturbation, condition \textbf{C2} means that the regressor contains at least one {\emph{known}} element that is independent from the perturbation, requirement \textbf{C4} is the condition of exponential stability of the estimation error $\tilde \kappa \left( t \right)$. It is shown by the numerical experiments that these conditions can be met.

\end{document}